\documentclass[11pt]{article}

\usepackage[margin=1in]{geometry}
\usepackage{times}
\usepackage{latexsym}
\usepackage{amsmath}
\usepackage{amssymb}
\usepackage{microtype}
\usepackage{hyperref}
\usepackage{url}
\usepackage{graphicx}
\usepackage{booktabs}
\usepackage{natbib}
\usepackage{caption}
\usepackage{subcaption}
\usepackage{enumitem}
\usepackage{setspace}
\usepackage{titlesec}
\usepackage{abstract}
\usepackage{newtxtext, newtxmath}

\newcommand{\keywords}[1]{\par\noindent\textbf{Keywords:} #1}

\titleformat{\section}{\large\bfseries}{\thesection}{1em}{}
\titleformat{\subsection}{\normalsize\bfseries}{\thesubsection}{1em}{}
\titleformat{\subsubsection}{\normalsize\itshape}{\thesubsubsection}{1em}{}

\hypersetup{
    colorlinks=true,
    linkcolor=blue,
    citecolor=blue,
    urlcolor=blue
}

\title{\textbf{Analytics for Quality Assurance for Item Pools (AQuAP): Monitoring and Maintaining Item Bank Health in AI-Driven Assessment Systems}}

\author{
  Alina A.\ von Davier \quad Xiaowan Zhang \quad Yigal Attali \quad Yena Park \\
  Jacqueline Church \quad Andrew Runge \quad Geoff LaFlair \quad Alexander Tsigler\\[4pt]
  \textit{Duolingo} 
}

\date{}

\begin{document}

\maketitle

\begin{abstract}
The large-scale digitization of educational assessment has made the continuous oversight of item banks both essential and complex. This paper presents \textbf{Analytics for Quality Assurance for Item Pools (AQuAP)}—a dashboard environment for monitoring item quality and item bank health. AQuAP supports  the operational implementation of the large scale item generation procedures for high-stakes tests as included in the Item Factory, a framework for automated and human-supported test development. The paper describes AQuAP in relationship with the process of item development, outlines the broader metric framework for item-pool quality assurance, and highlights the Effective Bank Size (EBS) as one central indicator of pool vitality. EBS quantifies how many independent test sessions can be constructed before content repetition occurs and, when coupled with exposure and usage metrics, provides insight into item bank security, diversity, and efficiency. We further introduce bank-health metrics---maximum exposure, maximum conditional exposure, adjusted effective bank size, and the rarely-administered fraction---that extend this picture of item utilization. AQuAP illustrates how operational analytics can translate psychometric concepts into quality assurance tools for high-volume, AI-enabled testing programs. This work is illustrated with the Duolingo English Test (DET) processes. 
\end{abstract}

\section{Introduction}

In digital assessments that operate continuously and at scale, the quality and security of item banks must be monitored with the same rigor applied to psychometric modeling or score interpretation. Automated item generation and adaptive delivery—technologies increasingly common in digital testing—create both opportunities for innovation and vulnerabilities related to item reuse, exposure, and parameter drift.

In the recent years, many assessment programs  developed multiple technology-based solutions to typical test development issues. The typical areas of interest for quality assurance in a high-stakes AI-based test are (a) the quality of the items and item bank, (b) the quality of the scores, and (c) the quality of the individual testing process. Specifically, Duolingo created an AI-empowered human-in-the-loop approach to automatically generate items for the Duolingo English Test (DET) and to monitor the quality of the item bank. This approach was implemented in the Item Factory \citep{vonDavier2024} and in the Analytics for Quality Assurance of Item Pools (AQuAP), respectively. A similar platform for monitoring the quality of the test scores, Analytics for Quality Assurance for Assessment \citep[AQuAA;][]{liao2022,liao2022psychometric} was also created and a third platform, Analytics for Quality Assurance for each Test Taker (AQuATT; in progress) to monitor for any discrepancies in the responses of a test taker within the test administration is being developed. The role of these three platforms and automatic tools is to continuously, effectively, and efficiently monitor the quality of the test from multiple perspectives, and to inform assessment experts of potential problems on an ongoing basis.

The Item Factory was conceived as a comprehensive framework for managing the life cycle of assessment content through automation and human oversight. It integrates human subject-matter expert (SME) design, artificial intelligence (AI) for item generation, automatic filtering for bias detection, SME review, and microservices for piloting and calibration. In conjunction with this production environment, AQuAP supports analytic oversight of the expanding item pool.

In this paper we describe the AQUAP platform that can be of interest to any high-stakes assessment program with a large-scale item generation process and a rapid or continuous administration mode that impacts the item exposure. AQuAP was conceived as a living monitoring system, automatically updated with data from the operational assessment environment.

\section{Context: From the Item Factory to Continuous Monitoring}

\begin{figure}[h]
  \centering
\includegraphics[width=\linewidth]{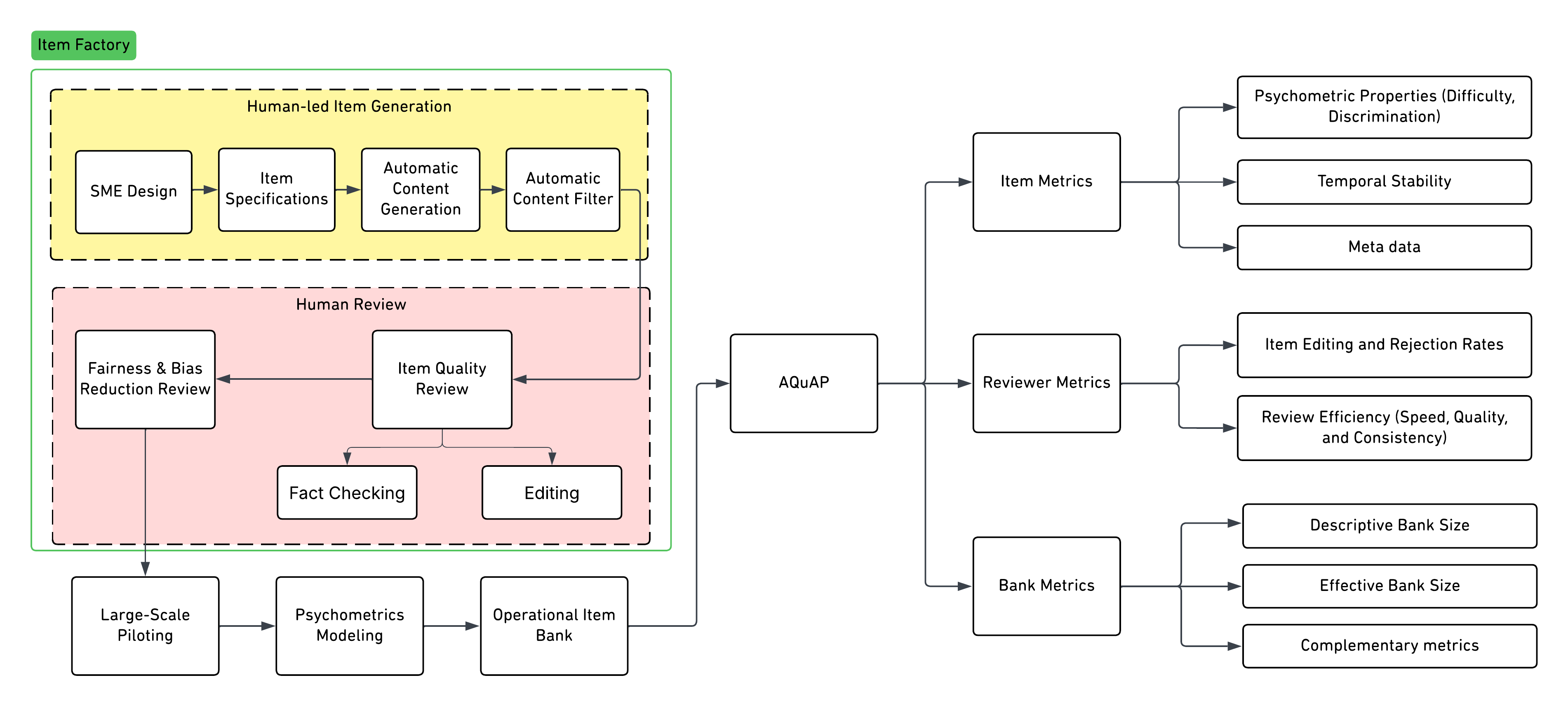}
  \caption{Illustration of DET's Item Factory and AQuAP.}
  \label{fig:factory}
\end{figure}

The Item Factory is a framework for AI-empowered, human-in-the-loop test development that adapts design ideas from smart-manufacturing systems—modularity, flexibility, and scalability—to the domain of assessment development. The model envisions a collaborative, data-driven ecosystem where machines and human experts share responsibilities. \citet{vonDavier2024} describes the Item Factory in detail. Here we provide a brief overview based on the item factory developed for the DET (illustrated in Figure~\ref{fig:factory}).

Items are first designed by SMEs; then they are machine-generated based on content needs and predefined specifications articulated by SMEs. These items undergo an automated content review for initial quality screening that checks length, grammatical and spelling accuracy, and other basic requirements. Items that pass the automated review are subsequently submitted to human experts for a two-phase review process:

\begin{itemize}[noitemsep]
  \item \textbf{Phase I: Item Quality Review (IQR).} Reviewers ensure that items effectively target the intended construct, demonstrate coherence, and maintain linguistic accuracy. Items deemed low quality are revised or discarded depending on the severity of the issue, and then undergo fact checking.
  \item \textbf{Phase II: Fairness and Bias (FAB) Review.} Reviewers ensure all test takers have an equal opportunity to demonstrate their language skills regardless of their backgrounds by identifying and eliminating potentially biased or culturally sensitive content. Failing items are discarded.
\end{itemize}

\noindent Items that successfully pass all review stages are then piloted, and their psychometric properties (e.g., difficulty and discrimination) are empirically estimated. Items demonstrating satisfactory performance in the pilot stage are added to the operational item bank, where their parameters and performance metrics are monitored as part of item-pool quality assurance.

The Item Factory approach to test development greatly accelerates the rate of item production. However, it also multiplies the need for dynamic oversight. With thousands of items being created, static quality checks prevalent in traditional item bank management systems are no longer sufficient. Continuous analytics must be implemented to provide the mechanism for ongoing verification.

AQuAP is a monitoring dashboard designed to support evaluation of the overall health of the item bank. The current Sigma \citep{sigma2025} implementation provides views of pool size by item type, item age, difficulty and exposure, weighted Effective Bank Size (EBS), response time and response length, input length, daily session volume, and item review status. The metric framework described in the remainder of this paper reflects the broader goals of the test's item-pool quality assurance; incorporation of additional analyses into the dashboard is ongoing work. Rather than treating quality control as an end-stage task, AQuAP is embedded directly into the life cycle of the Item Factory.

\section{Design Principles of AQuAP}

AQuAP is  a living monitoring system, automatically updated with data from the operational assessment administration. Its architecture follows three guiding ideas:

\begin{enumerate}[noitemsep]
  \item \textit{Automated data collection and visualization.} Information on item performance, exposure, and reviewer activity is ingested continuously and presented in dashboards and visual summaries.
  \item \textit{Human interpretation and intervention.} Experts remain responsible for evaluating anomalies and making strategic decisions such as item retirement or pool expansion.
  \item \textit{Integration of psychometric, operational, and editorial metrics.} AQuAP links classical psychometric indicators with workflow measures—bringing together item statistics, reviewer behavior, and item life-cycle data.
\end{enumerate}

The result is an evolving view of the test's operational health, accessible through a visual interface. One can also operate a separate alerting system---for example, notifying experts of changes in response time via Slack---whose integration with AQuAP is planned.

\section{Core Metrics}

AQuAP consolidates diverse metrics into three broad categories that together reflect the state of the item ecosystem. Below we outline the metrics that inform the item-pool quality assurance.

\subsection{Item-Level Metrics}

\paragraph{{\textit{Psychometric properties.}}}
Essential item-level psychometric properties include difficulty, discrimination, and model-fit indices derived from ongoing calibration. For multiple-choice items, distractor performance is also monitored to minimize the possibility of double keys \citep{orlando2000,sinharay2003}.

\paragraph{\textit{Temporal stability.}}
Item age and possible degradation or shifts in item functioning are tracked via temporal stability measures, such as trends in item difficulty parameters or changes in response time over time. Statistical process control tools such as CUSUM can be used to check for changes in the difficulty of an item over time, presumably due to item exposure. Declining item fit or abnormal response times can signal compromised content or overexposure \citep{lee2014security}.

\paragraph{\textit{Metadata-related metrics.}}
A wide range of metadata characterizes item content (e.g., domain, topic, rhetorical purpose, CEFR level) and associated assets (e.g., length tags for texts; character, accent, and speed tags for audio files; domain tags for images). These metadata support ongoing evaluation of the construct representativeness of the item bank. For instance, the domain coverage of a particular task type can be assessed by examining item counts by domain, and further insights can be gained by comparing item difficulty and discrimination parameters across domains.

\subsection{Reviewer and Process Metrics}

\paragraph{Item editing and rejection rates.}
The number of items that pass or fail each of the two human review processes, IQR and FAB, is tracked. For the IQR, it additionally records the number of items that (a) pass without editing and (b) pass after editing. For items with multiple components (e.g., multiple-choice items consisting of a prompt, stem, and distractors), it further documents which specific components are edited. Large amounts of post-generation human editing or high rejection frequencies can indicate systemic weaknesses in automated generation. These data can also be analyzed at the individual reviewer level as an additional measure of rater efficiency. For example, reviewers who pass all items may be flagged for additional training or calibration.

\paragraph{\textit{Reviewer efficiency.}}
Reviewer speed, quality, and consistency are tracked to evaluate reviewer efficiency. Review speed is measured by the number of items reviewed within a given time frame (e.g., weekly, monthly, or annually). These metrics are tracked longitudinally to identify trends and determine whether reviewers become more efficient over time. Before beginning the review process, each reviewer undergoes extensive training and calibration. Calibrated items are periodically embedded within the review pool to assess reviewer accuracy and monitor intra-reviewer reliability. Each item undergoes double review for FAB. Measures of inter-rater consistency (e.g., Quadratic Weighted Kappa) are used to ensure standardization and monitor inter-reviewer reliability.

\subsection{Bank-Level Metrics}

\paragraph{Descriptive bank size.}
At the macro level, descriptive bank size measures summarize the health of entire item pools, including the number of items in the bank and breakdowns by task type.

\paragraph{Effective Bank Size (EBS).}
By accounting for item administration, EBS provides a more informative aggregate indicator for bank size. Details about EBS are found in Section~\ref{sec:ebs}.

\section{Bank level metrics}
\label{sec:bankhealth}

In this section, we define the notions of Effective Bank Size (Section~\ref{sec:ebs}), and its adjusted version (Section~\ref{sec:aebs}) as the main aggregate metrics of the health of the item bank. Additionally, in Section~\ref{sec:complementary_metrics} we define complementary metrics aimed at identifying localized inefficiencies that aggregate measures may obscure. Section~\ref{sec:metrics_interplay} then explains how those metrics can be used for monitoring, while Section~\ref{sec:metrics_interplay} explains how they can be estimated and controlled.

\subsection{The Effective Bank Size (EBS) Metric}
\label{sec:ebs}

EBS expresses how many complete, unique test sessions can be produced before any item repeats. For each item type, it is calculated as:

\begin{equation}
  \text{EBS}_{\text{type}} = \frac{\text{Number of items of that type in the bank}}{\text{Number of such items used per session}}.
  \label{eq:ebs}
\end{equation}

The EBS can theoretically range from just above 1 to arbitrarily large values, 
though in practice it is bounded by the total number of items in the bank and 
the session design. A value of 1 indicates that every item in the bank is used 
in every session, meaning any two consecutive sessions would share all items---a 
situation of minimal content security. Values between 1 and 10 suggest a 
constrained bank where item overlap across sessions remains high and replenishment 
should be prioritized. Values in the range of 10 to 50 reflect a moderately 
healthy bank with acceptable content security for most operational settings. 
Values above 50 indicate a robust bank in which the probability of item 
repetition across sessions is low, providing strong protection against item 
exposure and memorization. As a general guideline, a minimum EBS of 20 is 
recommended for high-stakes assessments, while lower-stakes or pilot contexts 
may tolerate values closer to 10.

A higher value means that test takers are less likely to encounter repeated items across sessions, implying stronger content security and greater variety.

To produce a single, interpretable key performance indicator for the entire test, AQuAP computes a weighted aggregate across item types. Each weight corresponds to the proportion of testing time devoted to that item type:

\begin{equation}
  \text{EBS}_{\text{aggregate}} = \sum_{i} w_i \times \text{EBS}_i,
\end{equation}

\noindent where $w_i$ reflects the relative time allocation. This weighted value allows test developers and psychometricians to communicate the overall robustness of the item bank in one concise number while preserving sensitivity to the contribution of each item type.

\subsubsection{Practical Applications of EBS}

Within AQuAP, EBS serves several complementary purposes:

\begin{enumerate}[noitemsep]
  \item \textit{Monitoring bank health.} Drops in EBS over time indicate that certain item types are being depleted faster than replenished.
  \item \textit{Evaluating maintenance actions.} Whenever new items are added or obsolete ones removed, AQuAP recomputes EBS to quantify the change's impact.
  \item \textit{Communicating with stakeholders.} Because it summarizes complex patterns in a single number, EBS is used as a headline KPI in dashboards and reports.
  \item \textit{Informing generation priorities.} Item types with lower EBS values can be flagged for accelerated generation or additional review.
\end{enumerate}

\noindent In operational use, the EBS curve shows characteristic dips after the release of a new test version, followed by recovery as new content is introduced. These dynamics allow managers to anticipate when item pools will require expansion.

\subsection{Extending EBS: Effective Bank Use (EBU) and Adjusted Effective Bank Size (AEBS)}
\label{sec:aebs}

EBS does not account for the adaptive administration algorithm, which inherently causes some items to be selected far more frequently than others. To adjust for this, AQuAP's designers proposed a refinement called \textbf{Effective Bank Use (EBU)}—a scaling factor between 0 and 1 that reflects how evenly items are utilized:

\begin{equation}
  \text{AEBS} = \text{EBS} \times \text{EBU}.
  \label{eq:ebs_adjusted}
\end{equation}

In Equation~\eqref{eq:ebs_adjusted} AEBS stands for \textbf{Adjusted Effective Bank Size}: the 
exposure-corrected bank size, representing a more realistic estimate of the 
number of unique sessions that can be delivered when item selection is uneven. 
The term $\text{EBS}$ retains its original meaning from Equation~\ref{eq:ebs}: 
the theoretical maximum number of non-repeating sessions. EBU is derived from the entropy $H$ of the empirical item-administration distribution:

\begin{equation}
  \text{EBU} = \frac{1}{n}\exp\!\bigl(H(p_1, p_2, \ldots, p_n)\bigr),
\end{equation}

\noindent where $p_i$ is the empirical proportion of administration allocated to item $i$, and $H$ is the Shannon entropy. AEBS generalizes the EBS framework by grounding the measure in information theory, and is directly comparable across banks of different sizes. A higher AEBS indicates more uniform use of the bank, and the scaling factor $\text{EBU} \in [0, 1]$ captures the degree of 
balance in item utilization across the bank. When $\text{EBU} = 1$, all items 
are drawn with equal frequency, and AEBS reduces to 
EBS, recovering the baseline estimate. As item exposure becomes 
increasingly concentrated in a small subset of items, EBU approaches 
$0$, and AEBS shrinks proportionally, reflecting the 
loss of effective capacity. For example, an $\text{EBS} = 40$ paired with an 
$\text{EBU} = 0.25$ yields an $\text{AEBS} = 10$, indicating 
that despite a nominally large bank, uneven exposure reduces operational 
capacity to the equivalent of only $10$ unique sessions. Practitioners should 
therefore monitor both quantities jointly: a high EBS paired with a 
low EBU may give a misleading impression of bank robustness, whereas 
$\text{AEBS}$ surfaces the true security risk.

\noindent A low EBU value signals imbalanced use of the bank, where a small subset of items dominates test exposure. Incorporating EBU allows for a more realistic assessment of test security and supports algorithmic improvements in item selection strategies. An even further extension of EBU that is a function of ability may describe data from an adaptive test better and is currently being considered.

\subsection{Complementary Metrics}
\label{sec:complementary_metrics}
Item exposure and bank health can be characterized through three complementary metrics that together yield a more complete picture of item utilization patterns~\citep{attali2024}. Whereas EBS and EBU summarize overall bank capacity, these metrics reported here (that were also included in~\cite{sharpnack2025s2a3}), surface localized or structural inefficiencies that aggregate measures may obscure.

\paragraph{1.\ \textit{Maximum Exposure.}}
The highest marginal administration probability across all items in the bank. A ceiling of approximately 0.20 is a common standard for high-stakes tests. When any single item approaches or exceeds this threshold, it signals that the item is at elevated risk of compromise through repeated exposure in a broad population.

\paragraph{2.\ Maximum Conditional Exposure (MCE).}
The highest exposure probability within any ability stratum. MCE catches items that are overexposed to a specific subpopulation even when their marginal exposure appears acceptable. In adaptive testing, items near the population mean may have modest marginal exposure yet dominate administrations within a particular ability band, a risk that MCE makes explicit.

\paragraph{3.\ Rarely-Administered Fraction.}
The proportion of items whose exposure rate is less than 3\% of the exposure rate that would be expected under uniform (non-adaptive) administration. This metric tracks whether the tail of the bank is being neglected. Items with very low exposure rates may represent poorly calibrated content, content misaligned with the examinee population, or simply items that have not yet had sufficient opportunity to circulate. Monitoring this fraction helps identify items that should either be promoted into regular rotation or flagged for review and potential retirement.

\subsection{Interplay Among Bank-Health Metrics}
\label{sec:metrics_interplay}
\begin{sloppypar}
All of these metrics complement one another. High maximum exposure combined with high rarely-administered fraction indicates a severely skewed distribution in which a handful of items dominate test delivery while most items lie dormant. Low AEBS confirms this imbalance in information-theoretic terms. High MCE without elevated marginal exposure suggests that the item-selection algorithm is concentrating specific items within ability strata, a problem invisible to marginal analysis alone.
\end{sloppypar}
As in all quality assurance activities, one is looking for convergent information. For example, an unexpected drop in reviewer efficiency may slow the rate at which new items enter the pool, which in turn depresses EBS and AEBS while pushing maximum exposure upward. All metrics are evaluated together by human experts on a regular basis and interpreted in light of the broader test environment.

\subsection{Estimation and Control of Bank-Health Metrics}
\label{sec:metric_control}
Estimating and controlling bank-health metrics raises two related challenges, discussed in greater detail in \cite{sharpnack2025s2a3}. First, several metrics are not directly observable from raw administration counts alone. For example, computing MCE requires conditioning exposure rates on test-taker proficiency, which must itself be estimated. In large item banks, marginal exposure probabilities are also small for most items, so reliable estimation of AEBS and the rarely-administered fraction may require substantial administration data.

Second, bank-health objectives must be balanced against measurement reliability. A policy that always administers the most informative available item may maximize information locally, but it can also concentrate exposure on a small subset of items and leave much of the bank unused. Conversely, policies that enforce more uniform exposure may reduce the information available for ability estimation. Effective item administration therefore requires explicit control of this trade-off.

Both challenges are addressed through simulation. For each significant change to the item bank, test simulations are run to predict bank-health metrics before deployment and to tune the adaptive administration algorithm accordingly. The optimization targets an appropriate balance between AEBS and test reliability, while treating maximum exposure, maximum conditional exposure, and the rarely-administered fraction as constraints. The simulation reproduces the administration logic and uses IRT modeling to represent the population of test takers.

\section{Integration with Computational Psychometrics}

AQuAP operates within the computational psychometrics framework \citep{vonDavier2021}, which merges traditional measurement models with machine-learning approaches. 
\begin{sloppypar}
The broader computational psychometrics environment uses a modular design in which analytic components---IRT calibration, exposure monitoring, fairness auditing---function independently but communicate through a centralized data layer, ensuring scalability and rapid iteration.
\end{sloppypar}
\section{Visualization and Alerting}

AQuAP translates analytics into interactive visualizations accessible through a cloud-based dashboard. Examples follow.

AQuAP is currently hosted in Sigma (Sigma Computing, 2025), a cloud-based data visualization platform that provides interactive tables and visualizations. As part of an effort to create a more unified ecosystem with other quality assurance systems, including AQuAA, the developer has been evaluating alternative platforms for AQuAP. During this process, several options were considered, including Hex (Hex Technologies Inc., 2025), Quarto Dashboards (Allaire et al., 2025), Dash (Plotly Technologies Inc., 2025), and Shiny (Posit Software, PBC, 2025).
Dash was ultimately selected because of its open-source nature and low cost, ease of deployment, support for deep linking to specific views, modular visualization architecture, and robust capabilities for parameter segmentation and filtering.

\subsection{Example Visualizations from AQuAP}

\begin{figure}[h]
  \centering
  \includegraphics[width=\linewidth]{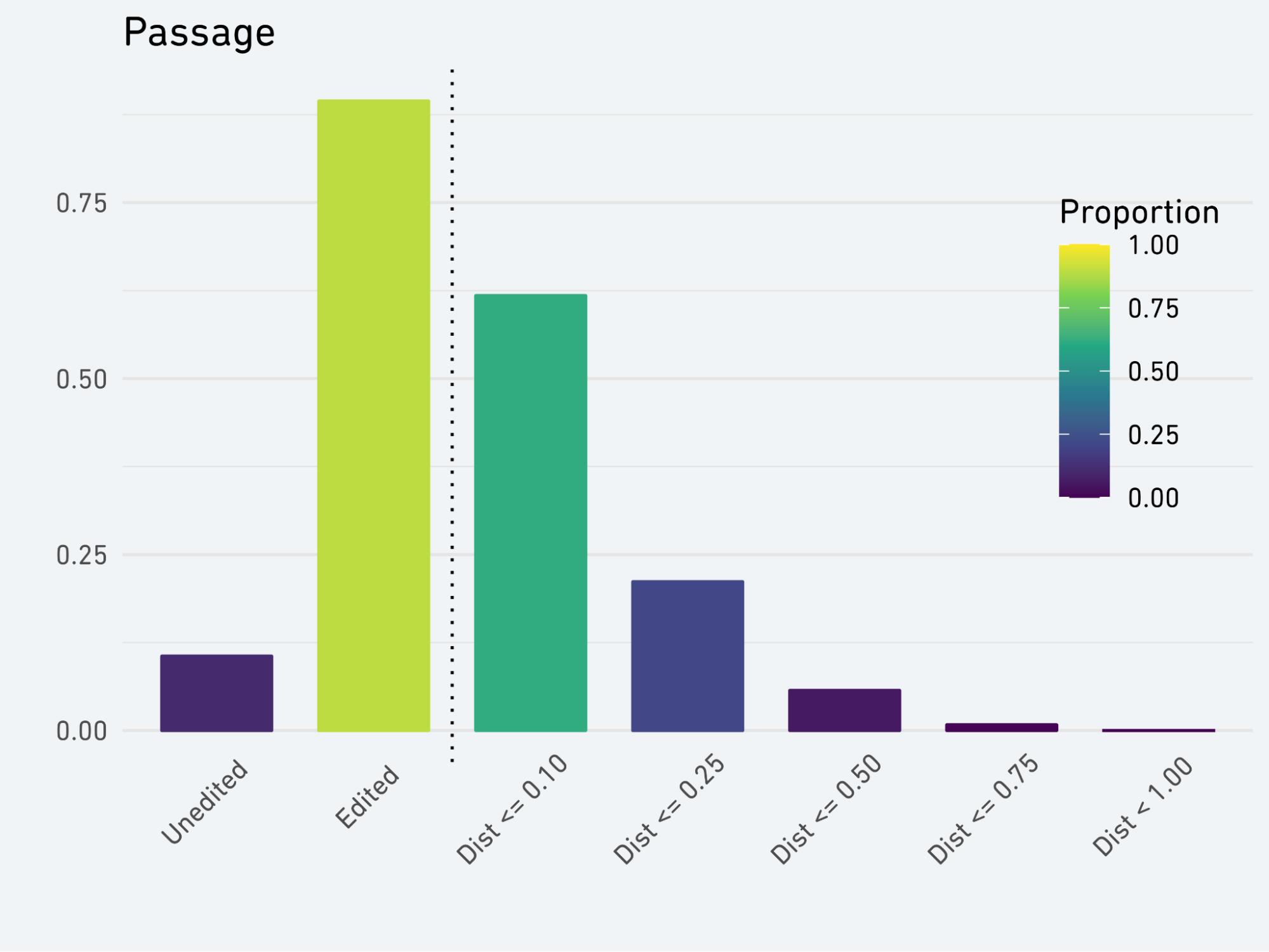}
  \caption{Passage-level edit counts and binned edit distance for interactive reading passages generated in 2023.}
  \label{fig:ebs}
\end{figure}

Figure~\ref{fig:ebs} illustrates the increase in EBS over a period of time for one specific item type, the c-test. A c-test is a type of language assessment item where within a paragraph parts of words are removed (according to rules), and test takers must complete the word to reconstruct the text.

\begin{figure}[h]
  \centering
  \includegraphics[width=\linewidth]{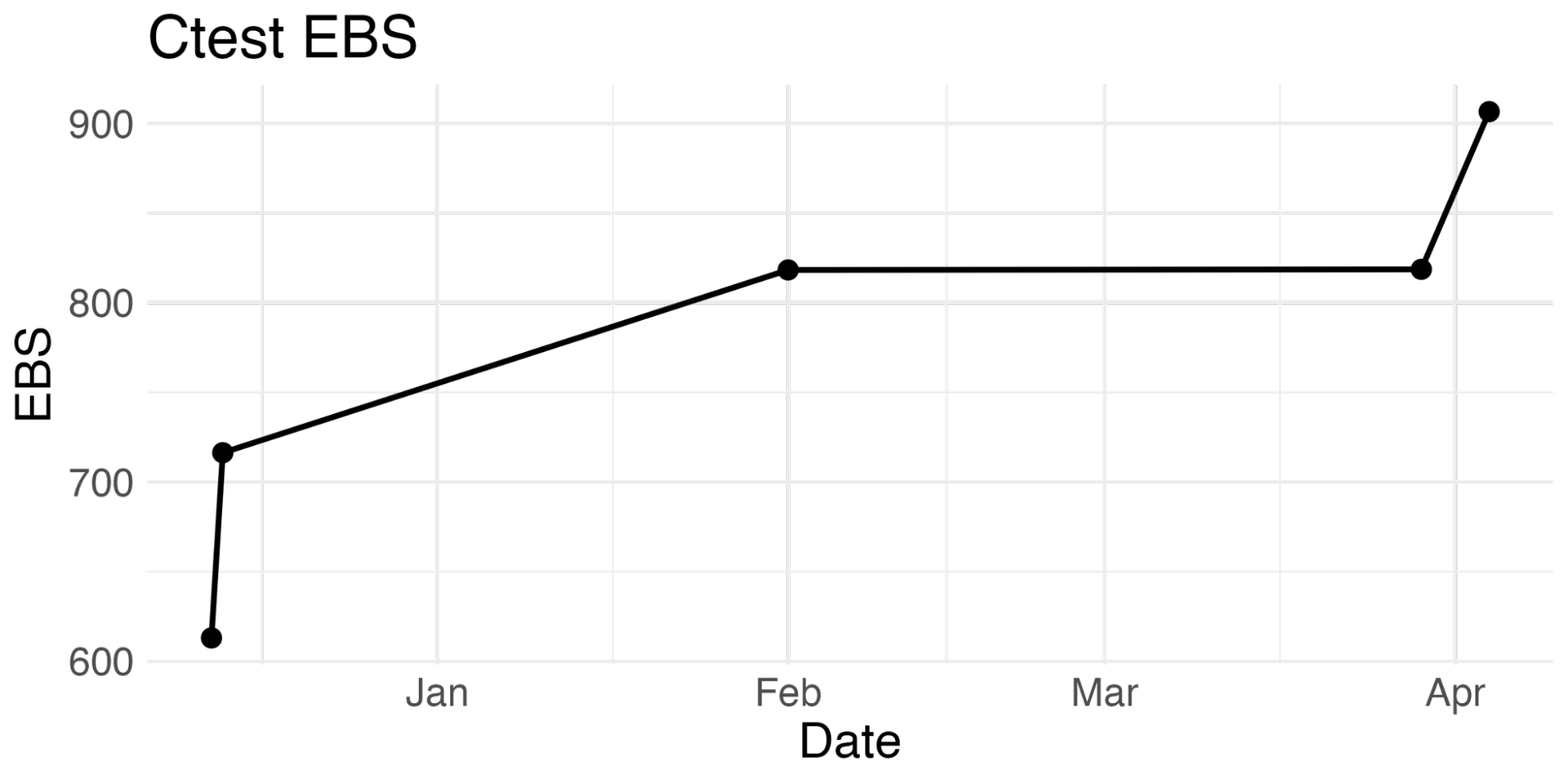}
  \caption{Effective bank size for c-test items over a period of eight months.}
  \label{fig:exposure}
\end{figure}

\begin{figure}[h]
  \centering
  \includegraphics[width=\linewidth]{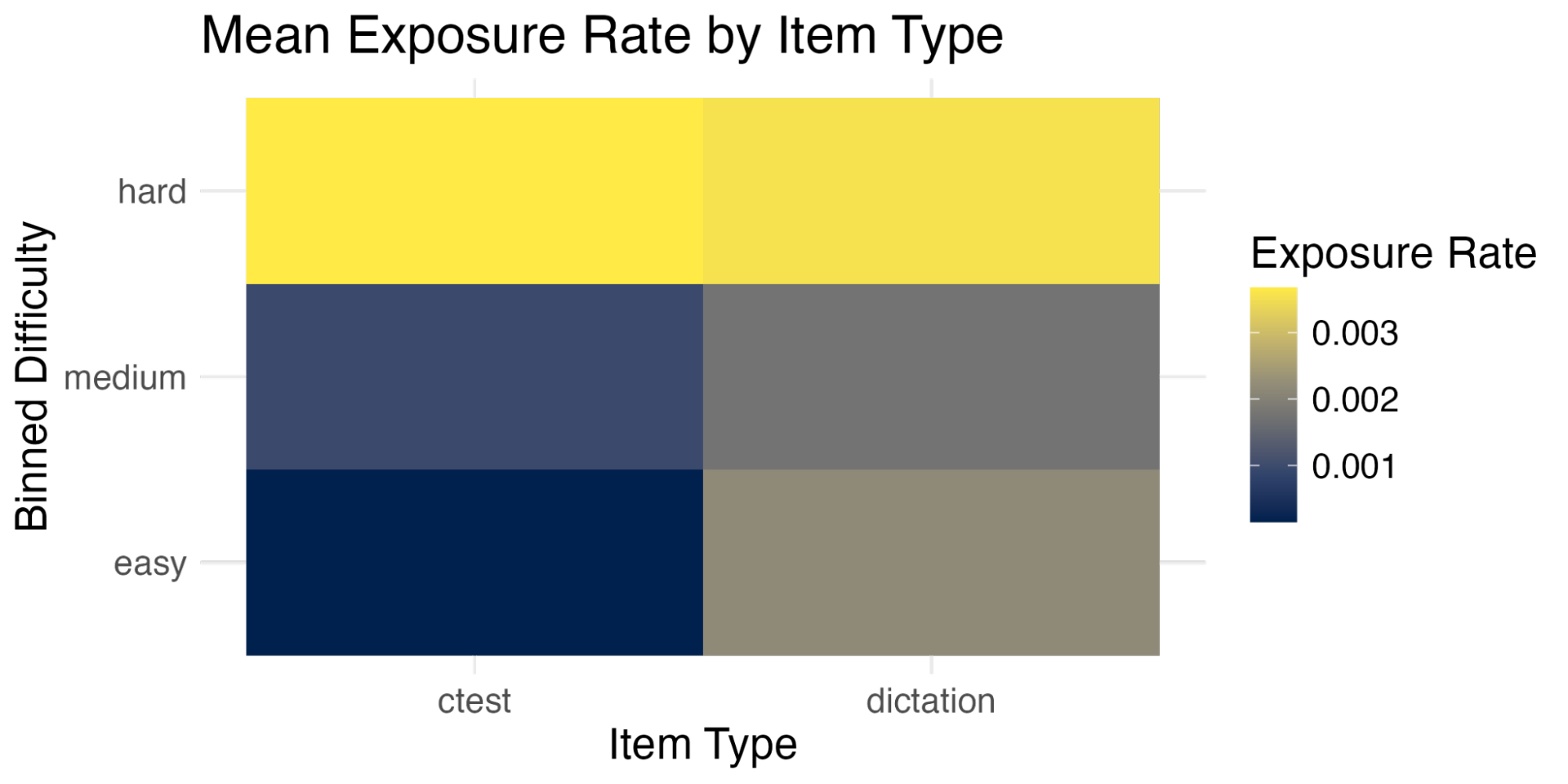}
  \caption{Mean exposure rate by binned difficulty over a rolling 30-day period.}
  \label{fig:edits}
\end{figure}

Figure~\ref{fig:exposure} illustrates exposure rate by difficulty for two item types over a rolling 30-day period. Figure~\ref{fig:edits} shows edit counts and edit distance for reading passages generated with earlier large language models.

\section{Role of AQuAP in the Item Factory Ecosystem}

Within the broader Item Factory (Figure~\ref{fig:factory}), AQuAP functions as a monitoring component connecting all stages of test development:

\begin{itemize}[noitemsep]
  \item \textit{Item Generation.} Confirms that automatically produced items cover the desired range of difficulty and content.
  \item \textit{Item Review.} Tracks reviewer reliability and throughput.
  \item \textit{Piloting and Calibration.} Observes the stability of item statistics in early use.
  \item \textit{Operational Testing.} Monitors live performance, security, and fairness indicators.
\end{itemize}

This continuous flow of data embodies the notion of quality by design—ensuring that every stage of item production and deployment feeds into an evidence-based monitoring cycle.

\section{Ethical and Operational Dimensions}

Embedding analytics in an assessment system must be done responsibly. Duolingo's Responsible AI Standards \citep{burstein2023} provide a framework for transparency and ethical oversight. Automated indicators such as EBS and AEBS are not treated as self-sufficient; they require expert interpretation to avoid decisions that could inadvertently affect fairness or construct representation.

Moreover, maintaining bank ``health'' also implies maintaining content diversity. A high EBS alone is not sufficient; the underlying material in the bank must represent balanced linguistic and cultural content. Therefore, AQuAP's quantitative outputs are reviewed in conjunction with qualitative audits conducted by human experts.

\section{Future Enhancements}
\label{sec:future}

Planned developments for AQuAP include migration to an in-house dashboard platform and automating additional item-pool analyses currently performed separately, among them:

\begin{itemize}[noitemsep]
  \item Full integration of EBU, AEBS, MCE, and the rarely-administered fraction;
  \item Discrimination, model-fit, distractor, and metadata breakdown views;
  \item Reviewer throughput, consistency, and component-level edit tracking;
  \item Integration of alerting with the dashboard;
  \item Predictive models that estimate when each item type will reach a critical exposure threshold;
  \item Automated anomaly detection for psychometric drift or irregular response patterns;
  \item Unified fairness dashboards that merge exposure metrics with differential-item-functioning analyses; and
  \item Extension to additional assessment products, allowing organization-wide monitoring.
\end{itemize}

These developments will extend AQuAP toward a predictive decision-support system for assessment quality management.

\section{Conclusion}

AQuAP demonstrates how principles of intelligent automation and operational analytics can be applied to maintain validity and fairness in large-scale, AI-enabled assessments. By quantifying item bank health through metrics such as EBS, EBU, AEBS, maximum exposure, maximum conditional exposure, and the rarely-administered fraction, AQuAP provides a growing foundation for monitoring test quality, security, and sustainability.

\begin{sloppypar}
As assessments move toward data-rich ecosystems, systems like AQuAP will become essential infrastructure---linking assessment science with responsible AI and engineering practices to ensure that educational measurement remains rigorous, equitable, and adaptive. AQuAP, its implementation, and this document will continue to evolve as the needs of AI-based assessment change.
\end{sloppypar}
\section*{Declaration on Generative AI}
During the preparation of this work, the author(s) used ChatGPT Plus in order to summarize information, improve grammar and edit, and reformat the references. Claude Sonnet 4.6 Enterprise was used to format the document to Latex/Overleaf. After using these tool, the author(s) reviewed and edited the content as needed and take(s) full responsibility for the publication’s content. 
\bibliographystyle{ACM-Reference-Format}
\bibliography{references_aquap}

@article{orlando2000,
  author    = {Orlando, M. and Thissen, D.},
  title     = {Likelihood-based item-fit indices for dichotomous item response theory models},
  journal   = {Applied Psychological Measurement},
  year      = {2000},
  volume    = {24},
  number    = {1},
  pages     = {50--64},
  doi       = {10.1177/01466216000241003}
}

@techreport{sinharay2003,
  author      = {Sinharay, S. and Johnson, M. S.},
  title       = {Simulation studies applying posterior predictive model checking for assessing fit of the common item response theory models},
  institution = {Educational Testing Service},
  year        = {2003},
  number      = {ETS RR-03-28},
  address     = {Princeton, NJ}
}

@incollection{lee2014security,
  author    = {Lee, Y.-H. and Lewis, C. and von Davier, A. A.},
  title     = {Test security and quality control for multistage tests},
  booktitle = {Computerized Multistage Testing: Theory and Applications},
  editor    = {Yan, D. and von Davier, A. A. and Lewis, C.},
  publisher = {Chapman \& Hall},
  address   = {London, UK},
  year      = {2014},
  pages     = {285--300}
}

@incollection{vonDavier2024,
  author    = {von Davier, A. A. and Attali, Y. and Runge, A. and Church, J. and Park, Y. and LaFlair, G.},
  title     = {The item factory: Intelligent automation in support of test development at scale},
  booktitle = {Machine Learning, Natural Language Processing, and Psychometrics},
  editor    = {Jiao, H. and Lissitz, R. W.},
  publisher = {Emerald Publishing},
  address   = {Leeds, UK},
  year      = {2024},
  pages     = {1--26}
}

@book{vonDavier2021,
  author    = {von Davier, A. A. and Mislevy, R. J. and Hao, J.},
  title     = {Computational Psychometrics: New Methodologies for a New Generation of Digital Learning and Assessment},
  year      = {2021},
  publisher = {Springer},
  address   = {New York, NY}
}

@techreport{attali2024,
  author      = {Attali, Y. and Church, J. and Park, Y.},
  title       = {Effective bank size metric ({WIP})},
  institution = {Duolingo},
  year        = {2024}
}

@techreport{burstein2023,
  author      = {Burstein, J.},
  title       = {Responsible {AI} Standards},
  institution = {Duolingo},
  year        = {2023},
  url         = {https://go.duolingo.com/ResponsibleAI}
}

@article{liao2022,
  author  = {Liao, M. and Attali, Y. and Lockwood, J. R. and von Davier, A. A.},
  title   = {Maintaining and monitoring quality of a continuously administered digital assessment},
  journal = {Frontiers in Education},
  year    = {2022},
  volume  = {7},
  pages   = {857496},
  doi     = {10.3389/feduc.2022.857496}
}

@incollection{liao2022psychometric,
  author    = {Liao, M. and Attali, Y. and von Davier, A. A. and Lockwood, J. R.},
  title     = {Quality Assurance in Digital-First Assessments},
  booktitle = {Quantitative Psychology: The 86th Annual Meeting of the Psychometric Society, Virtual, 2021},
  publisher = {Springer International Publishing},
  address   = {Cham},
  year      = {2022},
  pages     = {265--276}
}

@software{sigma2025,
  author = {{Sigma Computing}},
  title  = {Sigma},
  year   = {2025},
  url    = {https://www.sigmacomputing.com/}
}

@misc{sharpnack2025s2a3,
      title={S2A3: Thompson Sampling and Stochastic Exposure Control for High-Stakes CATs}, 
      author={James Sharpnack and Alexander Tsigler and J. R. Lockwood and Steven Nydick and Alina A. von Davier},
      year={2026},
      eprint={2606.07364},
      archivePrefix={arXiv},
      primaryClass={stat.AP},
      url={https://arxiv.org/abs/2606.07364}, 
}

\end{document}